# Multipactor in High Power Radio-Frequency Systems for Nuclear Fusion

Julien Hillairet[1], Marc Goniche[1], Nicolas Fil[1] [2] [3], Mohamed Belhaj[2], Jérôme Puech[3]

[1] CEA

CEA, IFRM, 13108 Saint Paul-Lez-Durance, France

[2] ONERA

ONERA The French Aerospace Lab, 31055 France

[3] CNES

CNES, DSO/RF/HNO, 31401 Toulouse Cedex 9, France.

## INTRODUCTION

Magnetic confinement fusion researches is the most advanced technique to master nuclear fusion for energy production. One of the main requirements for achieving fusion is to heat the plasma particles to temperatures exceeding 100-200 million of degrees (10-20 keV). Electromagnetic waves in mega-watt range of power, from tens of MHz to hundreds of GHz, are launched by antennas located near the plasma periphery in order to increase the plasma temperature and extend its duration [1]. These antennas, made of copper or silver-coated stainless steels, are located in vacuum environments. The vacuum sealing with pressurized transmission lines is made with the help of ceramics such as alumina, aluminium nitride, beryllium oxide or diamond. All these components are subject to multipactor discharges which increases the electron population (via secondary emission and gas desorption) which in turn may ultimately lead to an avalanche effect and the development of a discharge even at pressures two order of magnitude below Paschen breakdown limit [2]. These discharges are generally considered detrimental since they can lead to detuned RF systems, limit the RF power transmission in the plasma and eventually damage RF sources or components. When not detected quickly enough, arcs can lead to surface erosion [3], dielectric components metallisation [4] or water/air leaks due to punctured components such as bellows [5] or vacuum windows [6–8]. In some cases however, multipactor-induced discharges can be desired for vacuum RF conditioning during short RF pulses and at moderate power [3,4,9,10]. Moreover, these RF systems are subjected to the high magnetic field environment of the experiments in the Tesla range. The presence of magnetic field affects the electron trajectories and thus the multipactor resonances. On tokamaks, the electron motion is strongly constrained around the magnetic field lines which reduce the effects of loss mechanisms such as diffusion and increases electron impact ionization and thus the build-up of glow discharges. Finally, at the difference of RF payloads in satellite, the antenna surfaces can be polluted during operation with particles resulting from the strong interaction of the energetic particles with the walls of the tokamak, which may alter the surface characteristics such as secondary emission. This paper reviews the work performed in the fusion research community on multipactor discharges for two kinds of RF systems with their practical implications on power delivery into the plasma. The Ion Cyclotron Resonance Heating (ICRH), which uses coaxial lines in the MHz range of frequency, is presented in the next section. The Lower Hybrid Current Drive (LHCD) systems, which uses rectangular waveguides in the GHz range of frequency is discussed in section 3.

## ION CYCLOTRON RESONANCE HEATING ANTENNAS

The purpose of the Ion Cyclotron Resonance Heating system is to transfer the energy of high power electromagnetic waves launched by an antenna located near the magnetized plasma edge to heat the plasma. For tokamak magnetic field of few Teslas, the wave frequency is chosen between 30-80 MHz in order for the waves to exchange energy with some specific ion species of the plasma through a resonant process. This is one of the most used radio-frequency heating scheme on existing tokamaks [11,12], with up to 22 MW at 42 MHz of RF power being coupled to the plasma of the JET tokamak during few seconds [13]. In the ITER tokamak under construction in Cadarache, France, the ICRH system is designed to couple 20 MW at 40-55MHz to the plasma in Continuous Wave conditions (3600 s)[14].

In such systems, the RF power is transmitted from the multi-MW RF sources (generally tetrodes) to antenna(s) with 50 or 30 Ohm rigid coaxial lines made of steel or copper. Since the antenna is located in the tokamak vacuum environment, vacuum feedthroughs are necessary. Since the plasma equivalent impedance varies in time and does not match the generator output impedance, only a fraction of the power reaching the antenna(s) is transmitted to the plasma. In order to avoid the rest of this RF power to be reflected backwards and damaging the sources, devices such as 3 dB hybrid couplers or tuneable matching systems are used to either dump the reflected power or to recirculate it toward the plasma. The mechanical design and the electric scheme of the WEST tokamak ICRH antenna are illustrated in Fig. 1.

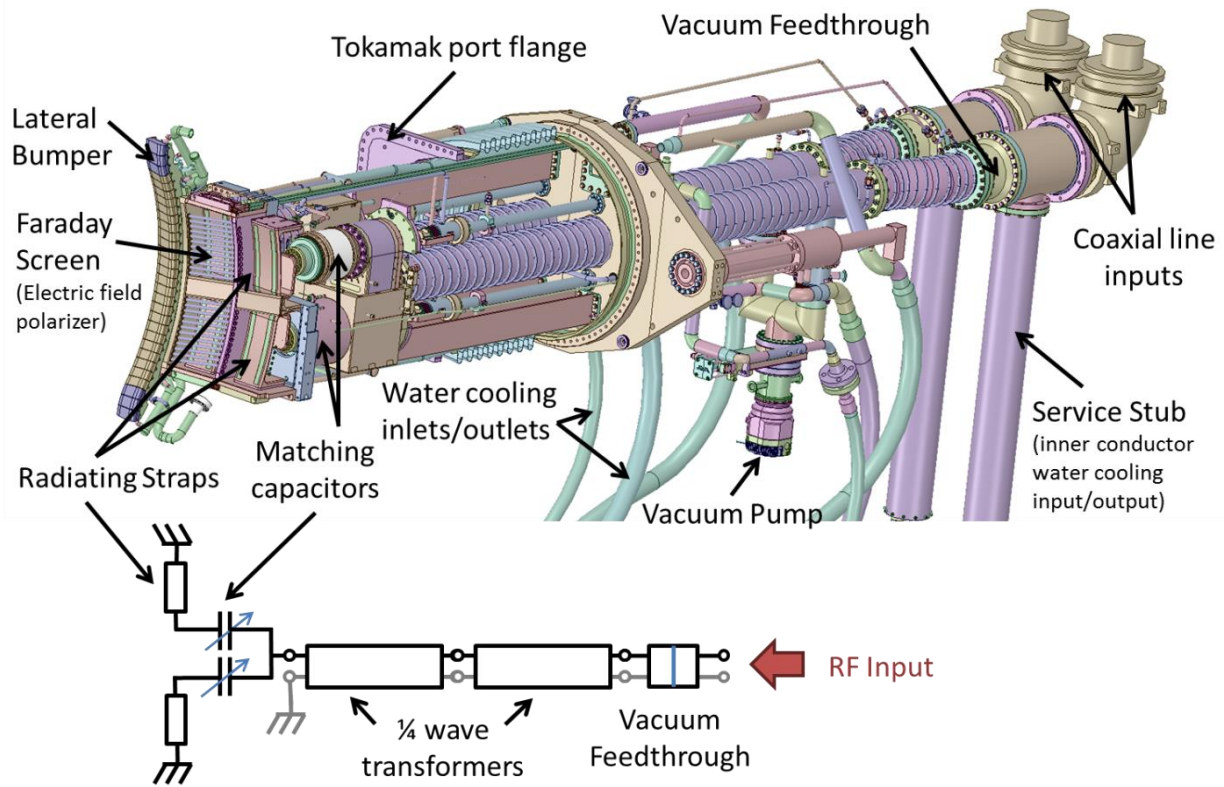

**Fig. 1. WEST Tokamak ICRH antenna mechanical and electrical design. The plasma (not illustrated) is facing the antenna at the left part of the Fig. .**

ICRH systems have general concerns with voltage limitations between metallic electrodes of coaxial lines or with dielectric elements of vacuum feedthroughs or inner conductor mechanical supports. Parasitic RF breakdowns can limit the power transmitted from the RF sources to the plasma. The processes which can be responsible for these voltage limits can be RF high-voltage arcs in vacuum or RF glow discharges (self-sustained or not, depending of the RF system configuration) [15]. Breakdowns appearing in the ICRH system can induce changes on the effective impedance which may detune the matching system and allows reflected power to return to the RF sources. Thus, reflection coefficient is the primary breakdown detection method. However, if the breakdown is close to a voltage node, the VSWR or reflection coefficient's phase may be almost unperturbed while increasing the apparent power coupling [16]. It is thus of great importance to detect these breakdowns as well as to switch-off sources to avoid damages. Additional detection techniques such as Sub-Harmonic Arc Detection (SHAD) [17], Scattering Matrix Arc Detection (SMAD) [18] or optical detection have been implemented or proposed in various systems to complete reflection coefficient method

[15,19,20]. High power breakdown experiments on dedicated test bed with large coaxial geometries relevant to ICRH have been setup in the fusion community, especially in order to understand breakdown conditions and improve arc detection methods [15,21]. Signature of multipactor breakdowns have been found to be low amplitude and wideband spectrum of excited frequencies at the opposite behaviour than high voltage arcs [19]. Voltage handling limits also arise in the antenna facing the plasma, which also reduce the ultimate antenna coupled power. Impurity production, increasing the neutral particles pressure, degrades the antenna voltage handling. Experimental works identified multipactor as the leading explanation for observed ICRF antenna neutral pressure limit [2].

In coaxial geometries, multipactor-induced breakdown disappear when the power level is above few kilowatts, which is a typical signature of multipactor resonance as expected from theory [22]. The later exact power level depends of the RF system such as Q factor and transmission line dimensions. The Fig. 2 (left) illustrates the multipactor power threshold versus the line impedance as calculated in a coaxial geometry with a fixed outer diameter at 50 MHz. Such breakdowns are known to happen when the power builds up during the start-up phase of the generators. For this reason, the RF rise time is generally set as short as possible, in order to exceed this power window as fast as possible ("push-through") and to be faster than the multipactor rise time [2,23,24]. Simulation of multipactor in a 30 Ohm coaxial line of dimensions ($D_{int}/D_{ext}$=140/230 mm) indicates that the number of initial seed electrons is doubled within 60-200 ns when RF power is below 2 kW (Fig. 2, right) In practice, the maximum current drawn by the generator or the RF power feedback systems limits this rise time to no more than typically 50 μs/kW, which is generally sufficient to avoid multipactor during power ramp-up.

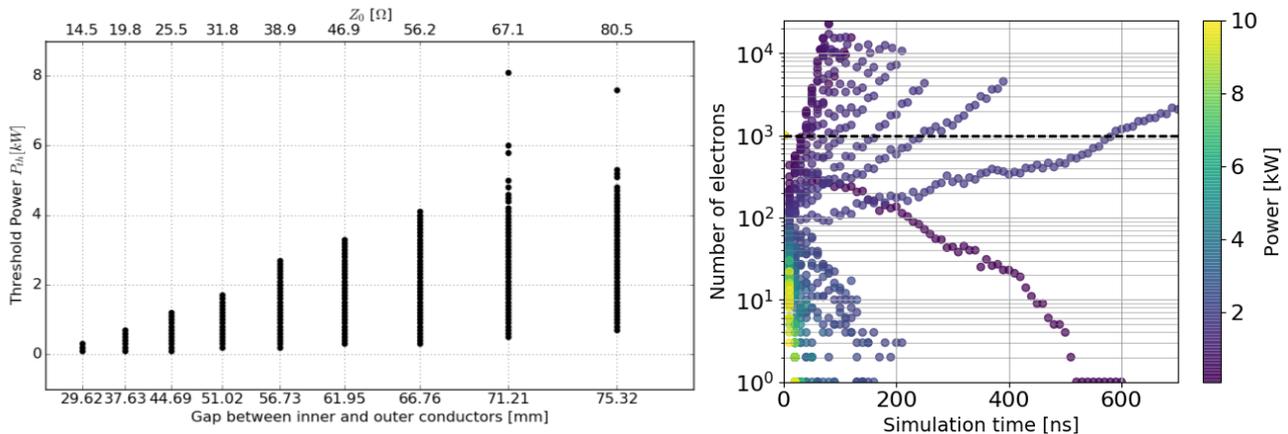

**Fig. 2. Left: Power Threshold in a coaxial line at 50 MHz calculated with Spark3D versus gap between inner and outer electrodes (with fixed outer diameter 153 mm) (Spark3D ECSS Copper SEY definition). Right: Number of electrons vs simulation time calculated with Spark3D in a 30 Ohm coaxial line (Spark3D ECSS Copper SEY definition). The horizontal dashed line represents the initial number of seed electrons (1000).**

In order to improve the tokamak vacuum conditions, the device temperature is raised between 120°C to 200°C during few tens of hours. This baking also reduces multipactor induced breakdowns by reducing the Secondary Emission Yield (SEY) of the conductor surfaces [25]. However, the direct baking of ICRH systems such as coaxial inner conductors or vacuum feedthroughs is often restricted because of their inaccessibility. Hence RF conditioning techniques are employed to reduce the adsorbed gas species trapped in the surface of the under vacuum conductor. Short RF pulses (few ms duration) are generally applied with power starting from hundreds of watts to kilowatts [4,24,26–28]. As the power increases, breakdown can be triggered by multipacting electrons [24,29]. When desired, these multipactor induced discharges can be beneficial as they help improving the general surface conditioning using multipactor induced glow discharges ("multipactor plasma") [24]. Since multipactor resonance conditions depend on the RF frequency and the geometry, changing the frequency allows sweeping the location of the triggered multipactor breakdowns.

When directed along the direction of propagation, a DC magnetic field can prevent the apparition of a multipactor discharge and its threshold depends of the local pressure [19]. However, when DC magnetic field is applied in the direction mainly transverse to the propagation direction, which is mostly the case in tokamaks, the neutral pressure at which a discharge is initiated in a ICRF antenna has been found to be reduced [30]. The magnetic force starts to compete with the electric ones from DC magnetic field of few $10^{-4}$ T. For stronger magnitude higher than 0.1 T, like the one founds in tokamaks, particles are constrained around the field lines with gyroradius lower than the gap between conductors. It has been found from simulations that the mean path length and time of flight are similar between unmagnetized and magnetized cases. So, the observed difference on ICRF antennas between these two cases is explained by the fact that magnetic field lowers onset voltages and minimum pressures at which glow breakdown occurs, as seen experimentally in [30]. This results in more severe pressure limitations in tokamaks.

**LOWER HYBRID CURRENT DRIVE ANTENNAS**

In a Tokamak device, the magnetic equilibrium is sustained by the current flowing in the plasma, which is magnetically induced by a solenoid located on the tokamak axis. This current can be maintained constant as long as the magnetic flux driving it is monotically varied, which leads to pulsed operations. In order to operate a tokamak in steady-state conditions, the plasma current must be generated by external means. The Lower Hybrid Current Drive (LH-CD) method has one of the highest current drive efficiency and is used in various tokamaks. The waves launched by antennas located nearby the plasma edge are launched with an asymmetric wave spectrum accelerating the electrons in a preferential direction around the torus. It results in a net current in the plasma which can replace partially or totally the induced current.

The LHCD antennas consist in arrays of phased rectangular waveguides under vacuum and have typically 100 waveguides for 3 MW antennas [31]. These antennas are made of several modules, each of them having one input waveguide and several narrower output waveguides facing the plasma (Fig. 3). In a so-called "multijunction" antenna, the input power is split into these narrower waveguides by thin E-plane walls. The phase shift between output waveguides is realized by sections of reduced waveguide heights which increase the wave phase velocity but also increase the electric field amplitude locally. The module input waveguide is connected via a vacuum RF window to a standard or oversized pressurized waveguide connected to the RF sources located meters away.

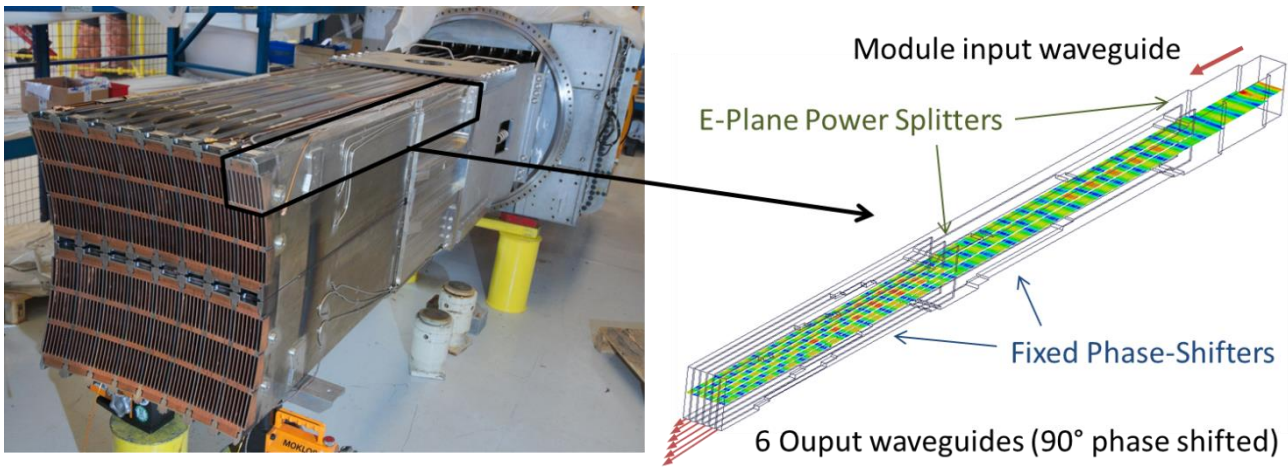

**Fig. 3. Left: One of the WEST Tokamak LHCD antenna. Right: detail of one multijunction module (1/48 of the antenna). The electric field magnitude is shown according to a color code.**

Multipactor has been identified from the beginning as one of the mechanism leading to reduce the power handling capability in rectangular waveguides in LH antennas [32,33]. It has been shown in [34] that the multipactor electric field breakdown threshold in rectangular waveguides scales almost linearly with the frequency. Converting to power threshold is not straightforward for complex LHCD antennas. Indeed, since the plasma always reflect 20 to 50% of the input power, this leads to standing waves patterns in the waveguide structure which locally increases the electric field for the same input power.

Near the front face of the antenna, a strong transverse DC magnetic field ($B_y \sim 1$ to 5 T) is mainly parallel to the RF electric field in the antenna's waveguides, with a strong negative gradient (1 to 2 T/m) along the waveguide longitudinal direction. Simulations showed a ~6% reduction of the multipactor threshold with respect to no magnetic field case for narrow waveguides (height<15 mm), as illustrated in Fig. 4 [35]. Moreover, strong resonance correlated with the increase of the multipactor order is numerically observed when the total field (in norm) is such that the electron cyclotron frequency equals the wave frequency. In this case, the multipactor threshold is decreased strongly for larger waveguides (-60%) than for narrower ones (-20%) [34]. However, such limitation is generally not observed experimentally. This is explained by the fact that the magnetic field is non-homogeneous in reality, which allows electrons to escape from the resonance zone.

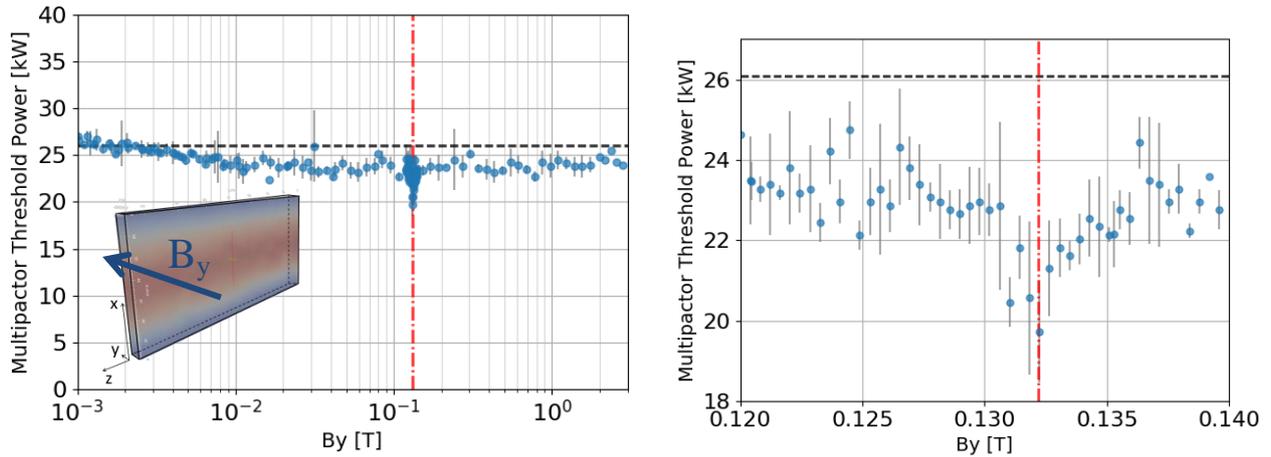

**Fig. 4. Multipactor power threshold for a 70x8mm waveguide at 3.7 GHz versus DC magnetic field parallel to the electric field. Right: zoom for $B_y$ in [0.12-0.14T]. Black dashed line is the multipactor power threshold without magnetic field. Red dot-dashed line corresponds to cyclotron resonance magnetic field. SEY from Spark3D ECSS Copper.**

When a magnetic field is directed along the waveguide's large side direction, the calculated multipactor power threshold increases or decreases of as illustrated in Fig. 5. However, in the typical range of magnitude observed in tokamak ($B_x \sim 0.1$ to 0.3 T, corresponding to the grey shaded zone), no multipactor occurs. The addition of this $B_x$ field to a $B_y$ component has no effect on the threshold, unless $B_x \sim B_y$ where it is enhanced [35].

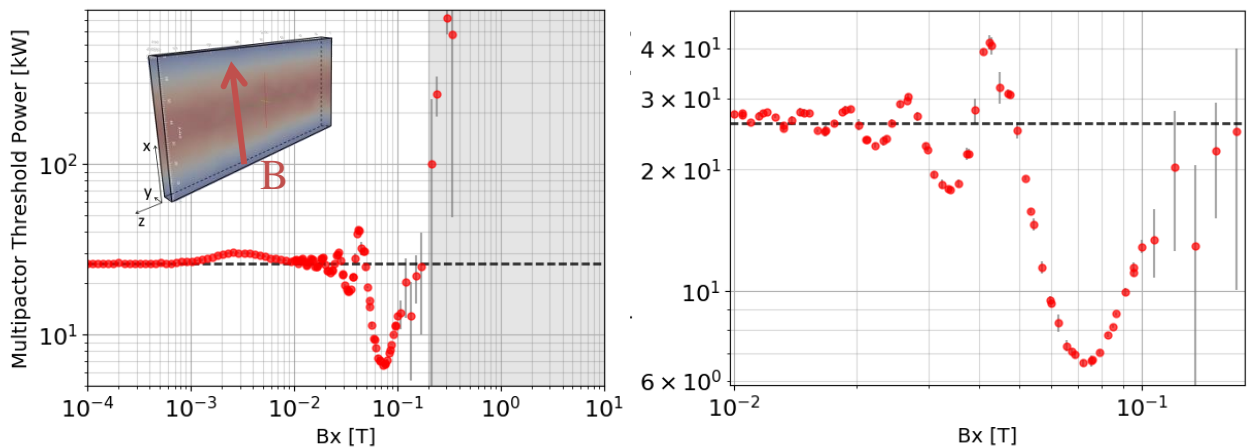

**Fig. 5. Multipactor power threshold for a 70x8mm waveguide at 3.7 GHz versus DC magnetic field parallel to the large side. Right: zoom for $B_x$ in [0.01-0.2T]. Black dashed line is the multipactor power threshold without magnetic field. SEY from Spark3D ECSS Copper.**

The vacuum feedthroughs are often made by an alumina ($Al_2O_3$) or beryllium oxide (BeO) ceramic slab brazed between two rectangular waveguides [8,36]. For steady-state systems, they are often located far from the mouth of launcher, outside of the vacuum vessel and the toroidal coils. The ceramic window can also be damaged by arc initiated near the antenna front face and travelling backwards to the RF source. Such arcs can deposit metallic particles on the ceramic surface, which would cause strong power handling limitations. Multipactoring induced breakdown on the ceramic is usually suppress by applying a thin (10 nm) TiN coating at its surface [37,38].

Similarly to ICRH, antennas are usually vacuum-baked to temperature from 120 to 200°C in order to reduce the surface contaminant desorption (C, O) [39] and the secondary electron emission yield [25], with the goal to reduce the apparition of breakdown. In addition, RF conditioning, which consists in applying short RF pulses repetitively on vacuum target (10 ms RF shots every 100 ms on Tore Supra), is also used in order to increase the power breakdown threshold [3,10,40].

**DISCUSSION ON FUTURE TOPICS**

Several aspects of multipactor-induced breakdown remain to be investigated in fusion RF systems. First of all, multipactor modelling work in the literature mostly concentrates on matched or low Voltage Standing Wave Ratio (VSWR) RF systems, which is the most frequent case in telecommunication satellite payloads. However, fusion RF antennas have to deal with a non-negligible part of the power reflected by the plasma. High VSWR can have an indirect effect on the multipactor threshold by peaking the electric field strength on several locations in the system and breakdown may appear at lower power level than initially assumed with matched models. Relationships between the plasma coupling properties with respect to the maximum power/electric field have to be established in order to help predicting the performances of the future large RF systems such as the ones of the ITER tokamak. Moreover, dielectrics like ceramics are used for the vacuum window materials in both ICRH and LHCD systems. These materials, mainly alumina or Beryllium oxide or eventually CVD diamond, have a high secondary emission yield and anti-multipactor treatments must be applied on them in order to allow high power handling. Future work will focus on the multipactor modelling of complex and multi-materials geometries with different secondary emission properties, with or without presence of inhomogeneous DC magnetic field in order to investigate the cyclotron resonance aspects in realistic environment.